# Observation of Meisser effect in Kapton Tapes

Jie Jiang[1,2†], Han Han[1,3†],Wei Xia[3], Yanfeng Guo[3], Yuanyuan Zhang[4], Haiping Fang[1], Long Yan[1*]

[1] Shanghai Institute of Applied Physics, Chinese Academy of Sciences, Shanghai 201800, China.

[2] University of Chinese Academy of Sciences, Beijing 100049, China.

[3] School of Physical Science and Technology, ShanghaiTech University, Shanghai 201210, China

[4] Key Laboratory of Polar Materials and Devices, Ministry of Education, East China Normal University, Shanghai 200241, China

[†] These authors contributed equally to this work

*Corresponding author. E-mail: yanlong@sinap.ac.cn (L. Y.)

**Abstract**

Composed of organic compounds, Kapton tape is regularly used as a non-conducting and non-magnetic material in cryogenic experiments. Here we report the discovery of Meisser effect at transition temperature (Tc) of ~ 3.8 K and ~ 8.0 K in Kapton tapes. New organic superconducting materials may be further explored in the tapes.

Discovering new superconductors and understanding their physics have been an important issue in condensed matter physics[1]. The discovery of organic superconductors[2] has expanded the superconducting field from inorganic materials to a wider area. Composed of various organic compounds, Kapton adhesive tapes are widely used in electronics manufacturing and scientific research experiments [3, 4]. Due to its low thermal conductivity, good dielectric quality, no magnetic elements contained, low outgassing rate, and good performance at low temperature [5], Kapton tapes are regularly used as a non-conducting and non-magnetic material in cryogenic experiments. It is intuitive to think that the Kapton tapes can not possibly have superconducting components.

Here we report on the detection of unexpected diamagnetic signals of Kapton tapes at low temperature. Two obvious diamagnetic signals associated with the Meissner effect are detected at the temperature of ~ 3.8 K and ~ 8.0 K, suggesting superconducting transitions in the Kapton tapes. To our knowledge, no superconducting materials have been reported in the components of the Kapton tapes. New organic superconducting materials may be further explored in this tape.

The Kapton tapes measured consist of polyimide (PI) films and some low temperature resistant adhesives [5]. The EDX measurement shows that the tapes only consist of C, N and O elements, and have not any metal element. A measurement of resistivity using four-point probe technique show that the surface resistivity of the tape is above 2 MΩ / sq. in room temperature. Magnetization measurements were performed with a SQUID magnetometer (MPMS3, Quantum Design).

The two fundamental features of superconductivity are Meissner effect and zero resistivity. However, due to preparation difficulty of the bulk organic superconductive crystals, the metallic state of the organic superconductors was hard to be realized even below Tc. Difficulties in detecting weak superconductivity from resistivity measurements can be overcome by using magnetic susceptibility experiments.

Fig. 1 shows the magnetic susceptibility ($\chi$) of a Kapton tape as a function of the temperature (T), under conditions of zero field cooling (ZFC) and field cooling (FC) at 10 Oe magnetic field. The $\chi$-T curve shows two sharp decreases at 3.3 K and 7.8 K in ZFC and FC measurements. Here, the two temperature points corresponding to the sharp decreases in $\chi$ are defined as the superconducting transition temperature $Tc_1$ and $Tc_2$. One can see that the superconducting transitions are very sharp and transition widths are less than 0.5 K, indicating two obvious diamagnetic signals associated with the Meissner effect [6]. This suggests that there exist superconducting transitions in the Kapton tapes.

Fig. 2 shows the $\chi$-T curves under different magnetic fields for the Kapton tape in the ZFC measurement. The magnetic susceptibility decreases gradually with increasing H at the range of 0.6 — 100 Oe (Fig. 2a). It should be noted that the superconducting transition temperatures $Tc_1$ and $Tc_2$ are both clearly observed below 100 Oe, while $Tc_2$ is also existed at 400 Oe (Fig. 2b). Fig. 3 exhibits the magnetic-field dependence of $Tc_1$ and $Tc_2$, in which the $H_0 \cdot (1\text{-}Tc^2/T_0^2)$ fitting curves are also drawn [7]. The first and second transition temperatures are evaluated to be 3.8 K and 8.0 K, respectively.

In order to eliminate the possibility of superconductor of PI, the magnetic susceptibility of a pure PI film was measured. The diamagnetic signals can not be observed above 2 K. Therefore, we speculate that the superconductivities come from the low temperature resistant adhesives in the Kapton tapes. The adhesives are usually made from some organic high-molecular polymer, such as siloxane organ.

In summary, our results indicate that the Kapton Tape is magnetic responsive at low temperature. The two diamagnetic signals associated with the Meissner effect are detected at the temperature of 3.8 K and 8.0 K, suggesting superconducting transition in the Kapton Tape. Our results suggest that its usage should be specially paid attention in the cryogenic experiments. The new superconducting organic materials and their superconductivity mechanism may be further explored.

**Acknowledgments:**

We thank Prof. Guosheng Shi and Prof. Chungang Duan for their constructive suggestions.

**Figures and Captions**

Figure 1

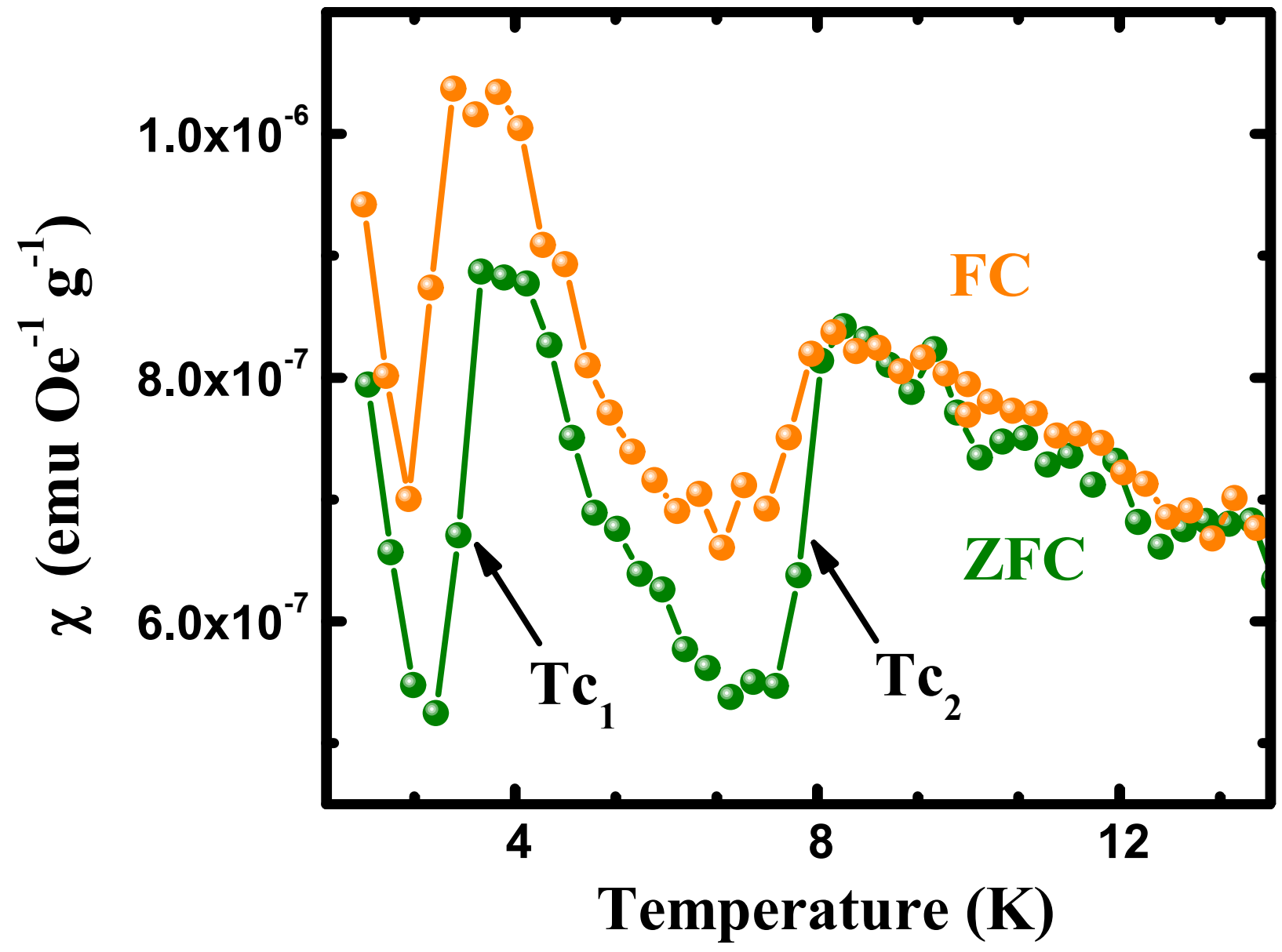

Fig. 1 Magnetic susceptibility of a Kapton tape as a function of temperature under conditions of zero field cooling (ZFC) and field cooling (FC) at 10 Oe magnetic field. The superconducting transition temperature $Tc_1$ and $Tc_2$ are indicated.

Figure 2

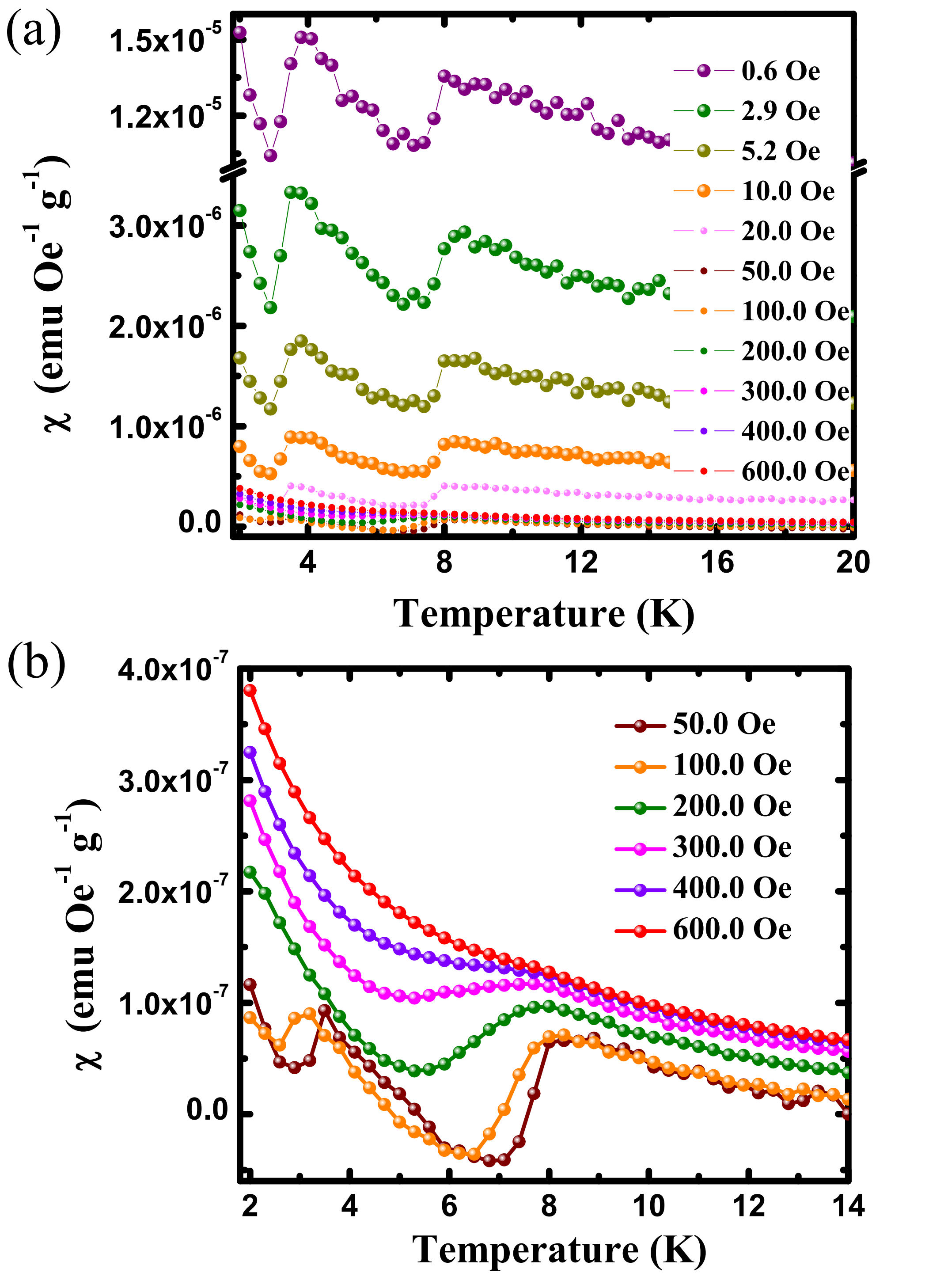

Fig. 2 Temperature dependence of χ under different magnetic fields for the Kapton tape in the ZFC measurement in the range of (a) 0.6 — 600 Oe and (b) 50 — 600 Oe

Figure 3

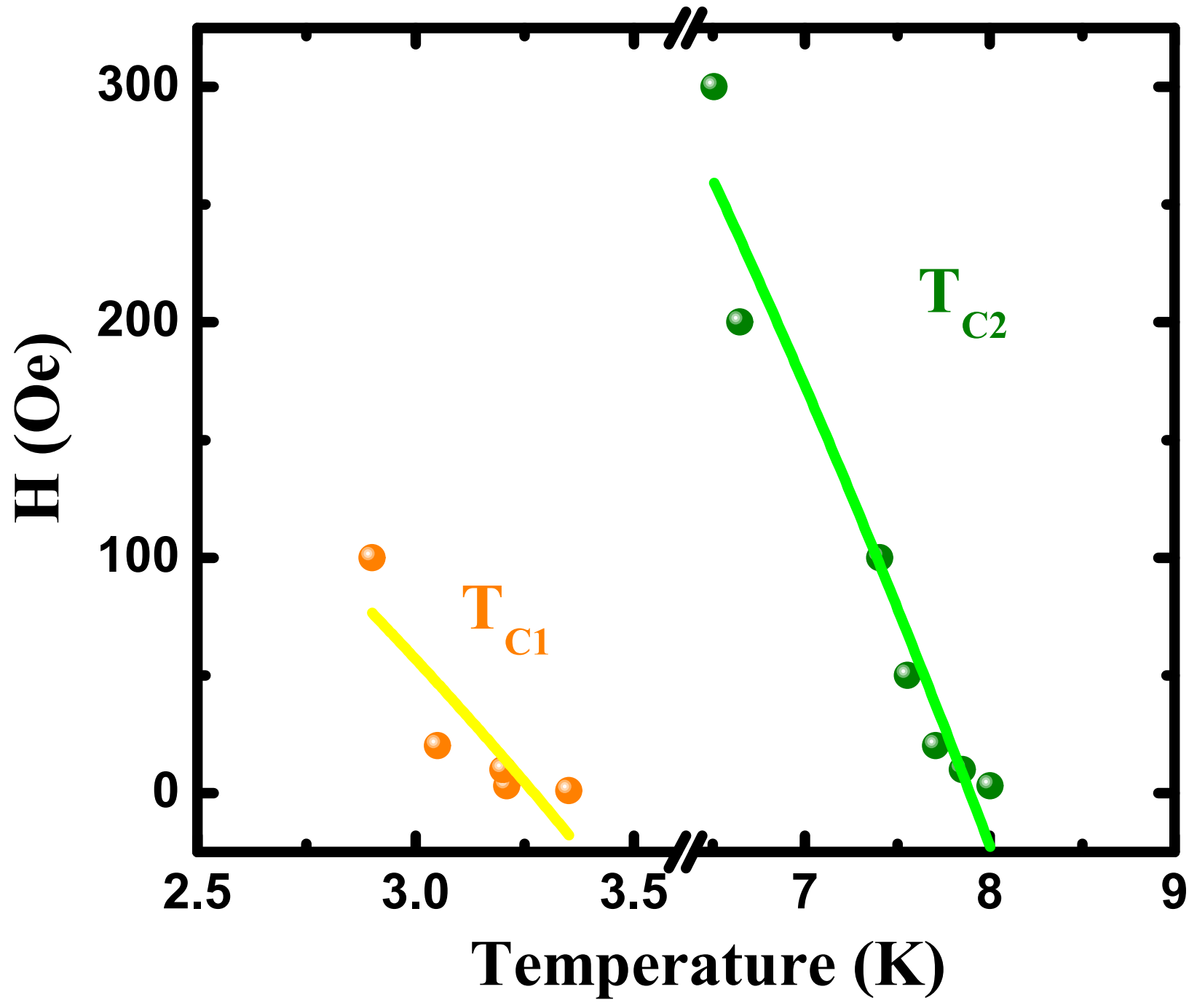

Fig. 3 Magnetic-field dependence of $Tc_1$ and $Tc_2$. Here, the $H_0 \cdot (1-Tc^2/T_0^2)$ fitting curves are also drawn. The first and second transition temperatures can be calculated 3.8 K and 8.0 K at zero magnetic field, respectively.